# Conformal TiO$_2$ aerogel-like films by plasma deposition: from omniphobic antireflective coatings to perovskite solar cells photoelectrodes

Jose M. Obrero,[a] Lidia Contreras-Bernal,[a] Francisco J. Aparicio,[a,b] Teresa C. Rojas,[a] Francisco J. Ferrer,[c] Noe Orozco,[a] Zineb Saghi,[d] Triana Czermak,[a] Jose M. Pedrosa,[e] Carmen López-Santos,[a,b] Kostya (Ken) Ostrikov,[f] Ana Borras,[a] Juan Ramón Sánchez-Valencia,[a]* Angel Barranco.[a]*

a) Nanotechnology on Surfaces and Plasma Laboratory, Materials Science Institute of Seville (CSIC-US), C/ Américo Vespucio 49, 41092, Seville, Spain.

b) Departamento de Física Aplicada I, Escuela Politécnica Superior, Universidad de Sevilla, Spain. c/ Virgen de Africa 4101

c) Centro Nacional de Aceleradores (CNA, CSIC-Universidad de Sevilla, Junta de Andalucía). Avda. Tomas Alba Edison 7, 4092, Sevilla, Spain and Dpt. Física Atómica Molecular y Nuclear, Facultad de Física, Universidad de Sevilla.

d) Univ. Grenoble Alpes, CEA, LETI, F-38000 Grenoble, France

e) Departamento de Sistemas Físicos, Químicos y Naturales. Universidad Pablo de Olavide, Ctra. Utrera Km. 1, 41013 Sevilla, Spain.

f) School of Chemistry and Physics and Centre for Materials Science, Queensland University of Technology (QUT), Brisbane, QLD 4000, Australia.

*corresponding authors: angel.barranco@csic.es, jrsanchez@icmse.csic.es

**Abstract**

The ability to control porosity in oxide thin films is one of the key factors that determine their properties. Despite the abundance of dry processes for the synthesis of oxide porous layers, the high porosity range is typically achieved by spin-coating-based wet chemical methods. Besides, special techniques such as supercritical drying are required to replace the pore liquid with air while maintaining the porous network. In this study, we propose a new method for the fabrication of ultra-porous titanium dioxide thin films at room or mild temperatures (T ≤ 120 °C) by the sequential process involving plasma deposition and etching. These films are conformal to the substrate topography even for high-aspect-ratio substrates and show percolated porosity values above 85% that are comparable to advanced aerogels. The films deposited at room temperature are amorphous. However, they become partly crystalline at slightly higher temperatures presenting a distribution of anatase clusters embedded in the sponge-like structure. Surprisingly, the porous structure remains after annealing the films at 450 °C in air, which increases the fraction of the embedded anatase nanocrystals. The films are antireflective, omniphobic, and photoactive becoming super-hydrophilic subjected to UV light irradiation The supported percolated nanoporous structure can be used as an electron-conducting electrode in perovskite solar cells. The properties of the cells depend on the aerogel film thickness reaching efficiencies close to those of commercial mesoporous anatase electrodes. This generic solvent-free synthesis is scalable and is applicable to ultra-high porous conformal oxides of different compositions with potential applications in photonics, optoelectronics, energy storage, and controlled wetting.

## 1.-Introduction

Porous nanostructured metals and metal-oxide films have attracted strong recent attention owing to their diverse emerging applications in catalysis, photonics, photovoltaics, electrodes for batteries, intelligent materials, energy harvesting, sensing, nanomembranes, and bio-interfaces.[1-4] To meet the requirements for most recent applications, it is essential to develop synthetic procedures for customized porous structures directly compatible with their fabrication on processable substrates such as transparent conductive electrodes (TCOs), flexible polymers, and on-device architectures such as interdigitated electrodes.[5-9] Tunable deposition of porous metal and metal oxide nanosystems including solution methods, electrochemical, electrospinning, as well as physical and chemical vapor deposition processes has recently emerged as a highly-topical area of research. [10-15]

Vacuum-based deposition methods are especially suited for the development of devices integrating functional coatings with advantages such as precise control over thickness and composition, uniformity, purity, versatility, and scalability to large-scale production as demonstrated by many examples from optoelectronics and microelectronics.[16-18] Properly designed, the vacuum-based processes can be energy efficient and produce a very low or negligible environmental impact. Porous structures have been developed by vacuum-based methods such as Oblique Angle Deposition (OAD) or Glancing Angle Deposition (GLAD) for many different technological areas.[14, 19] By tuning the deposition parameters, it is possible to obtain high surface area structures with anisotropic pore structure distribution and enhanced diffusion properties. For example, porosity control in $TiO_2$ films by OAD/GLAD[20] has been applied for the development of gas sensor[21,22], photonic structures[19, 23], microfluidic devices[24] and solar cell electrodes by liquid infiltration.[25] Other vacuum-based methods, such as plasma-enhanced chemical vapor deposition (PECVD) and magnetron sputtering, which typically produce compact and dense films, have also been applied for the development of porous oxide thin films for diverse applications, albeit to a lesser extent.[26-29]

Despite the many reports on the preparation of porous oxides by vacuum-based methods, the very high porosity range is typically achievable by wet chemistry methods. Aerogels are a class of nanostructured materials with ultra-high porosity (80-98%), extremely low density, and high surface area.[30] Aerogels are used in structural applications, as catalysts and catalytic supports, adsorption materials and thermal insulators.[31,32] Aerogels are usually prepared in a powder form by sol-gel methods combined with special drying methods such as supercritical drying to remove the solvent and fill the network with air avoiding the collapse of the original gel structure due to surface tension. Different wet methods such as spin-coating, dip-coating, and spray-coating have been proposed to obtain supported aerogel thin films.[33,34]

Titanium oxide films have been the subject of extensive research and application due to their remarkable properties. They have demonstrated utility as photocatalysts, electron transport layers in solar cells, optical materials due to their high refractive index and effective absorption of UVA radiation, gas sensors, components in biomaterial development for medical applications, surfaces with bactericidal properties and dielectric protective layers.[35-38] The properties of titanium dioxide films as band-gap position, crystallinity, and porosity are highly dependent on their preparation conditions.[39,40] $TiO_2$ aerogels were first prepared by Teichner in 1976.[39] Since then, they are usually prepared from Ti alkoxides and inorganic salts (e.g., TiCl$_4$), mostly for photocatalytic and thermal applications.[32]

In this work, we present a new approach for the synthesis of ultra-porous conformal TiO$_2$ dioxide films reaching the levels of high porosities and low densities characteristics of aerogel materials. The deposition process combines remote plasma polymerization of a Ti phthalocyanine precursor, to produce conformal homogeneous Ti-containing plasma polymers, and a plasma etching to remove the organic component of the polymer yielding highly porous TiO$_2$ films. The combined plasma polymer deposition and plasma etching process can be repeated successively to increase the thickness of the films increasing the overall porosity. Due to their structure, the TiO$_2$ aerogel-like films developed are antireflective and omniphobic and can be further thermally annealed to increase their crystallinity without collapsing the porous structure. Interestingly, despite the high porosity levels achieved and partial crystallinity, the films are photoactive. An initial feasibility study about the use of the TiO$_2$ aerogel-like films as electron transport layers in perovskite cells shows results comparable to standard commercial mesoporous titania coatings fabricated by sol-gel methods indicating the potentiality of this type of films for their integration in devices after a proper optimization. This technique is generic and can be applied to other metal precursors to synthesize aerogel-like conformal functional oxide nanocoatings. In addition, the plasma and vacuum processes employed are industrially scalable.

## 2.-Experimental

**Aerogel-like thin film synthesis**: The main precursor molecule in this study, Titanium (IV) phthalocyanine dichloride (TiPc), and additional precursors such as Titanyl phthalocyanine, Ti acetylacetonate, Silicon (IV) phthalocyanine dichloride, and Iron (II) phthalocyanine were purchased from Sigma-Aldrich and used as received.

Conformal metal-containing plasma polymer films were deposited by remote plasma-assisted vacuum deposition (RPAVD). Full details of the experimental setup can be found elsewhere.[42,43] In summary, the precursor molecules were sublimated in the downstream region of an electron cyclotron resonance microwave (ECR-MW) Ar plasma (150 W, 2.45 GHz) using a Knudsen cell placed at about 10 cm from the plasma. The substrates were at room temperature (RT) and placed facing the Knudsen cell and facing away from the plasma discharge. The growth of the polymer film is monitored in situ with a quartz crystal microbalance (QCM) placed next to the sample holder. Argon pressure was dosed by a calibrated mass flow controller to reach a deposition pressure of 2x10$^{-2}$ mbar. The base pressure of the reactor was <10$^{-6}$ mbar.

Silicon (100), and fused silica were used as substrates. Special substrates as supported nanowires on Si(100) for core@shell deposition and perovskite solar cell electrodes were prepared as specified below.

**Plasma etching treatments by ECR-MW**: The post-treatment of the plasma polymer films by plasma etching was carried with the samples facing an Ar/O$_2$ (1:1 mass flow) plasma discharge (2x10$^{-2}$ mbar, 350 W). The distance between the substrates and the glow discharge region was 8 cm. Substrate temperatures from RT to 120 °C and treatment durations from several minutes up to 30 minutes depending on the sample characteristics. Cycles of the plasma polymer deposition and plasma etching were conducted in the same reactor by rotating the sample holder and using the described experimental conditions.

**Synthesis of supported organic nanowires (ONWs) by Physical Vapor Deposition (PVD)**: $H_2$-Phthalocyanine (H2Pc) from Sigma-Aldrich, was used as received. The base pressure in the deposition system was $10^{-6}$ mbar. The sublimation of the molecule was carried out using a Knudsen cell, placed at 8 cm from the substrates, under $10^{-2}$ mbar of Ar flow, which was dosed by a calibrated mass flow controller. The growth rate and equivalent thickness of H2Pc NWs were monitored using quartz crystal microbalance (QCM), and the growth rate was adjusted to 0.3-0.4 Å/s setting a density in the QCM electronics of 0.5 g/cm$^3$. The substrate temperature was imposed at 170 ºC in a heatable sample holder connected to an electric current source, to induce the formation of the organic nanowires. Additional details about the synthesis of ONWs can be found elsewhere.[44-45]

**Preparation of perovskite solar cells.** Fluorine-doped tin oxide glass TEC 15 (FTO, XOP Glass, resistance 12-14 Ω/□) were used as substrates for the cells. The substrates were cleaned using an ultrasonic bath and following the sequence of solvents: Hellmanex solution (2:98 %V soap/water), deionized water, isopropanol, and acetone (15 min for each solvent). After the cleaning sequence, the substrates were dried using a nitrogen gas flow. Then, a UV/$O_3$ treatment was applied for 15 minutes. Thin compact layers (20 nm) of $TiO_2$ (c-$TiO_2$) were deposited on top of the FTO substrates by spray pyrolysis. The precursor of the compact layers consisted of 1 ml of titanium di-isopropoxide bis(acetylacetonate) solution (75% in 2-propanol, Sigma-Aldrich) in 14 ml of absolute ethanol. The precursor solution was sprayed over annealed FTO (400 ºC) using oxygen as carrier gas. The FTOs were kept at 450 ºC for 30 min to achieve the anatase phase. Once the substrates were cooled down, the ultra-porous $TiO_2$ layer was deposited as indicated above.

For the preparation of reference cells, a mesoporous solution was prepared by adding 1 mL of absolute ethanol to 150 mg of a commercial $TiO_2$ paste (Sigma-Aldrich, 18NRT). The mesoporous dispersion was left under stirring overnight. After that, 100 µL were deposited on c-$TiO_2$ samples by spin-coating at 4000 rpm for 10 s. After spinning, the samples were immediately placed on a hot plate at 100 ºC for 10 min. The mesoporous $TiO_2$ layer (m-$TiO_2$) was then sintered following a programmed temperature variation according to previous report.[44]

Perovskite layers were grown on the ultra-porous electrode following the method reported by Saliba et. al.[44] This is, a perovskite precursor solution consisting of (($FAPbI_3$)$_{83}$($MAPbBr_3$)$_{17}$ + 5% CsI) + 5% RbI was prepared by the addition of 1 M formamidinium lead triiodide (FAPbI3) and 1M methylammonium lead tribromide (MAPbBr3) solutions (5/1%V, respectively) both in 1:4 %V DMSO:DMF (dimethyl sulfoxide and N,N-dimethylformamide, respectively). A 5 %V of 1.7 M of CsI solution in DMSO was added to this solution, and then 5% V of RbI solution in 1:4 %V DMSO: DMF (0.2:99.8%mol, respectively). The perovskite film (RbCsMAFA) was deposited by two-step spin-coating: 1) 1000 rpm, 10 s; 2) 6000 rpm, 20 s. 15 s after the beginning of the second step, 200 µL of chlorobenzene was added as antisolvent. Immediately, the substrate was annealed at 100 °C for 45 min. Once the substrate was cooled down, a doped solution of Spiro-OMeTAD was deposited by spin-coating on top of the perovskite layer. This solution consisted of 70 mM 2,2,7,7-tetrakis[N,N-di(4-methoxyphenyl)amino]-9,9-spirobifluorene (Sigma-Aldrich) in chlorobenzene and the dopants Lithium bis(trifluoromethanesulfonyl)imide (LiTFSI, 520 mg/mL in acetonitrile), tris(2-(1H-pyrazol-1-yl)-4-tert-butylpyridine) cobalt(III)tris(bis(trifluoromethylsulfonyl)imide) and 4-tert-Butylpyridine in a molar ratio of 0.5, 0.03 and 3.3, respectively. The perovskite and Spiro-OMeTAD layers were deposited in a glovebox under very low humidity and oxygen conditions (< 0.1 ppm $O_2$ and <0.1 ppm $H_2O$).

Finally, gold contacts (thickness ~ 80 nm) were deposited on the devices by evaporation in high vacuum (<10$^{-6}$ mbar).

**Experimental characterization methods:** High-resolution scanning electron microscopy (SEM) images of the samples deposited on silicon wafers were obtained in a Hitachi S4800 microscope at an acceleration voltage of 2 kV. Cross-sectional views were obtained by cleaving the Si (100) substrates. Focused Ion Beam 3D analysis (FIB-3D) of selected samples was performed on a Zeiss crossbeam 550 FIB-SEM. Scanning TEM (STEM), High resolution transmission electron microscopy (HRTEM), and high-angel annular dark field (HAADF)-STEM images were acquired in a TALOS F200S microscope, working at 200KV and in a Tecnai G2F30 S-Twin STEM, working at 300KV, both with HAADF detector . Electron tomography was performed in a Titan Themis using a Fischione tomography holder.  DigitalMicrograh software has been used to analyze the HRTEM images and to get the digital diffraction patterns (DDP).

X-Ray Photoemission Spectroscopy (XPS) experiments were performed in a Phoibos 100 DLD x-ray spectrometer from SPECS. The spectra were collected in the pass energy constant mode at a value of 50 eV using magnesium and aluminum sources. C 1s signal at 284.8 eV was utilized for calibration of the binding energy in the spectra. The assignment of the BE to the different elements in the spectra corresponds to the data in XPS reference databases.[45]

UV-Vis transmission spectra of the samples deposited on fused silica slides were recorded using a Perkin Elmer spectrophotometer in the range from 190 to 2500 nm. Reflection spectra were recorded in a Cary 5000 from Agilent using a universal measurement accessory (UMA) in the range 190-2500 nm. Variable angle spectroscopic ellipsometry was carried out in a Woollam VASE ellipsometer.

Rutherford Backscattering Spectroscopy (RBS) and Nuclear Reaction Analysis (NRA) were carried in the 3MV Tandem Accelerator at the Centro Nacional de Aceleradores (US-CSIC, Sevilla, Spain). The RBS spectra were obtained using 2.0 MeV alpha particles and collecting the backscattered particles with a PIPS (passivated implanted planar silicon) particle detector at 165⁰. NRA was used to determine C, N, and O elements in the film from the $^{12}$C(d,p)13C, 14N(d,α1)12C y 16O(d,p1)17O nuclear reactions using deuterons of 1.0, 1.4 y 0.9 MeV, respectively. The spectra were obtained using a particle detector at a 150º collection angle in combination with a 13 μm thick Mylar filter to stop the retrodispersed particles. NRA and RBS spectra were simulated using the SIMNRA 6.0 code.[46] The film densities were determined from the combined RBS and NRA analyses and thickness values obtained from cross-sectional SEM micrographs of the films.[28]

Adsorption/desorption isotherms were measured with a QCM using quartz crystal sensors with deposited porous oxide layers following the procedure described in previous references.[28,49] For this analysis a set of samples were grown directly on quartz crystals in the plasma reactor. The isotherms were obtained by sequentially introducing varying amounts of water vapor into a closed chamber containing the QCM. Before the adsorption experiment, the samples were heated under vacuum at approximately 120 ºC to eliminate any condensed or adsorbed water in the films during their exposure to air. The total pore volume was estimated under the assumption that all pores were filled with water at the saturation pressure.

Atomic force microscopy (AFM) images were obtained using the tapping mode of a Nanotec microscope with Dulcinea electronics, and then analyzed with the WSxM software.[50]

The wetting behavior was studied by contact angle measurements of liquid droplets with Milli-Q water (2ul) and CH2I2 diiodomethane (1ul) by the static sessile droplet method on an OCA20 from DataPhysics Instruments GmbH. Contact angle values are presented as an average of 5 measurements. Fluorine-based chemical derivatization was carried out by exposing a previously plasma-activated surface to 1H,1H,2H,2H,2H-PFOTES vapor maintained in a thermal bath at 80 °C for 3 h after a prior vacuum pump. These conditions favor the reaction of the surface -OH groups with a perfluorinated silane.[51]

Photoactivation by UV irradiation in air were performed with a 175 W ASB-XE-175 xenon light source lamp.

### 3.- Results and discussion

*3.1. Development of ultra-porous aerogel-like $TiO_2$ films by plasma deposition and etching.*

In a previous work, we demonstrated that plasma oxidation of porphyrin and phthalocyanine sublimated films produced conformal nanostructured oxide surfaces.[9,42] Plasma oxidation effectively removes organic components from such films, producing oxide surface aggregates whose properties (i.e., oxidation state, degree of aggregation, and surface percolation) depend strongly on the nature of the metal cation and the thickness of the initial sublimated film. Therefore, porous nanostructured Zn, Co, Cu, oxide, and Pt metal surfaces were prepared by direct plasma oxidation of sublimated Zn(II), Co(II), Cu(II) and Pt(II) porphyrins or phthalocyanines.[44,52]

Herein, we present the evolved synthetic approach by exploiting the use of sacrificial polymeric films obtained by remote plasma-assisted vacuum deposition (RPAVD) of Ti(IV) phthalocyanine dichloride (hereafter TiPc plasma polymer). The RPAVD method gives rise to plasma polymer films by adjusting the plasma interactions with a sublimated functional precursor. The details of the methods have been discussed in previous works.[42,43,53] RPAVD films are continuous, conformal, homogeneous, insoluble, and crosslinked plasma polymers containing some fraction of unreacted precursor molecules depending on the synthetic conditions used. In this work, the RPAVD TiPc films are used as conformal sacrificial layers to be oxidized by interaction with oxygen plasma at low or room temperature (RT). The polymeric Ti-containing plasma polymer films act here as the precursor layers for Ti atoms that will eventually form $TiO_2$ films by reacting with the oxygen plasma species. In addition, the plasma etching, or plasma oxidation steps remove C and the remaining heteroatoms by forming volatile species that are pumped out of the chamber. The $TiO_2$ films are thus deposited by the iterative process shown in Figure 1a). In each of the synthetic cycles, a Ti-containing plasma polymer is deposited and then subjected to an oxygen plasma treatment to form a titanium oxide film. The plasma polymerization and oxidation cycle is repeated several times to increase the final layer thickness. As indicated in the schematic, each cycle comprises a conformal TiPc plasma polymerization (Step 1) and plasma etching (Step 2) steps. These cycles are required to fully oxidize the layers after the oxygen plasma treatment. As it will be shown below, thin layers of several tens of nm are fully oxidized at mild temperatures and plasma powers. However, the plasma polymer is not completely oxidized for thicker films under the same experimental conditions. Thus, the repeated cycling procedure is an essential requirement for the controlled growth of thicker and porous layers of $TiO_2$.

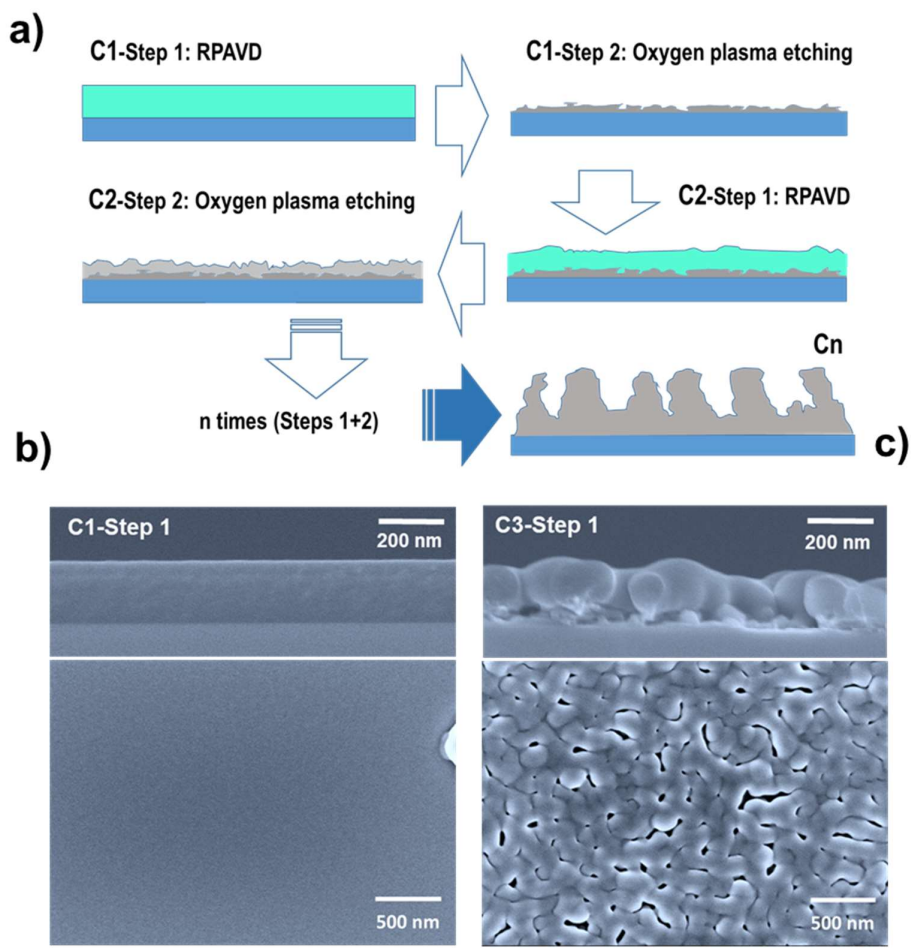

**Figure 1.- Sequential deposition procedure of porous oxide thin films by plasma polymerization and remote etching**. a) Schematic of the sequential plasma deposition (RPAVD) and plasma etching processes to obtain the ultraporous TiO$_2$ thin films. The thickness of the films increases as a function of the number of deposition (step 1) plus etching (step 2) cycles denoted as C1, C2,…,Cn. b) FESEM cross and planar micrographs of an RPAVD Ti-containing polymeric film on Si(100). This polymeric film deposition is the starting step for the synthesis of porous TiO$_2$ films. c) A RPAVD polymeric film deposited on top of a porous TiO$_2$ film grown on two previous cycles of plasma polymerization and plasma etching.

Figure 1b) shows the planar and cross-sectional films of a TiPc polymeric film deposited on Si(100) by RPAVD. The micrographs show that this film is continuous and homogeneous with a very smooth surface. AFM characterization of a set of polymeric films shows RMS surface roughness is < 0.5 nm for films with thicknesses in the range of 50-500 nm (data not shown); this is a common feature of polymer films synthesized by RPAVD.[42,43,53] These films exhibit intense absorption bands ascribed to embedded Ti phthalocyanine molecules (see Supporting Information S1). Apart from these localized molecular absorption features, the films are transparent in the visible (VIS) and near-infrared (NIR) range.

When the TiPc plasma polymer films are exposed to an oxygen plasma, oxygenated species from the plasma are incorporated to form nanostructured inorganic $TiO_2$ films. The C, N, and Cl elements from the initial polymeric films were removed as volatile species. The surface elemental and chemical analysis was conducted by X-ray Photoelectron Spectroscopy (XPS) on a sublimated TiPc film, a titanium-containing plasma polymer deposited using RPAVD, and three titanium dioxide ($TiO_2$) oxide films deposited through one, three, and five deposition and etching cycles (referred to as C1, C3, and C5, respectively) in Figure S2. The RPAVD process generated plasma polymers with a composition closely resembling the stoichiometry of the precursor molecule, exhibiting only a minimal oxygen enrichment of less than 10 atomic %. It is worth noting that such oxygen enrichment is characteristic of any plasma deposition process due to post-deposition reactions of the samples in air or by direct incorporation of residual oxygen species in the plasma.[42] After the plasma oxidation, the resulting $TiO_2$ films were found to be slightly over-stoichiometric (i.e., surface oxygen enrichment). Importantly, the films were analyzed in their as-deposited state without undergoing any surface cleaning procedures. A minor carbon content (~17%) was detected in the oxide films, particularly in C3 and C5. This carbon content and the excess of oxygen could be attributed to surface contamination (adsorbed hydrocarbon and water molecules and surface hydroxyl groups). Note that a small percentage of oxygen is also present on the surface of the precursor powder reference. However, a small percentage of carbon content from incomplete oxidation of the sacrificial polymer precursor cannot be avoided. Notably, no traces of nitrogen (N) or chlorine (Cl) from the precursor molecules were detected on the surfaces of the oxide films.

Figure 2a) shows the FESEM micrographs of the film shown in Figure 1b) after the plasma etching process. The figure shows how the initially homogenous polymeric films were converted into a porous structure presenting a distribution of nanometer-sized holes that are spread from the film surface to the Si substrate. These deposition and etching steps are repeated successively to increase the thickness of the resulting porous film (cycles Cn in Figure 1a)). Figures 2 a)-e) show that the film thickness increases with each cycle. The microstructure evolves from an open hole interconnected oxide to a highly open three-dimensional percolated porous structure. It can be noted that after every cycle, in the resulting $TiO_2$ films (Figure 2 b), the pores of the $TiO_2$ film increased in size with respect to that produced in the previous deposition-oxidation cycle. From the third cycle onwards, the film evolves into a more columnar-like structure, which is mainly formed by interconnected voids forming an open network where the empty space is predominant. This interconnected structure is preserved in cycles C4 and C5. The deposition and plasma treatment cycles shown are very effective in increasing the thickness of the films while maintaining the porous structure of the previous layers. This is possible because, in every first deposition step of each cycle, the TiPc plasma polymer conformally coats the open porous surface obtained in the previous cycle (as shown in Figure 1c) before the plasma oxidation. After plasma oxidation of the plasma polymer, the previous pore structure is preserved, and the new modified pore structure remains interconnected. Therefore, the conformal polymer coating also has the role of preventing the closure of the previously formed pores. This process is most likely due to the deposition of conformal polymeric films without diffusion inside the micropores. Microscopy measurements confirm that all pores of the film structure are hollow after the etching process, as will be seen below.

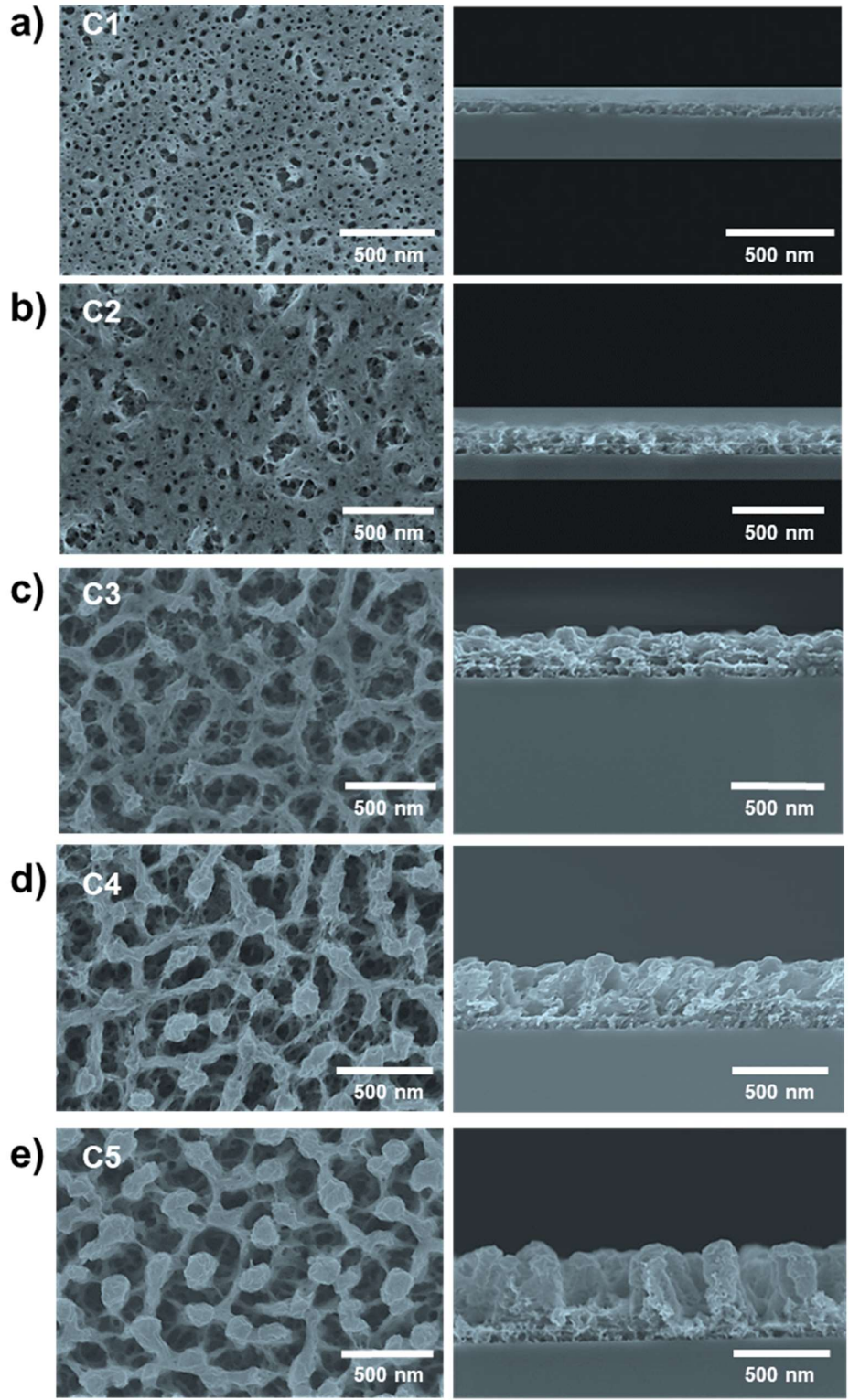

**Figure 2.- Characterization of film structure by scanning electron microscopy.** Top-view (left) and cross-sectional (right) FESEM images of films deposited on Si(100) using deposition-etching cycles. Panels a) to e) correspond to synthetic cycles C1 to C5, respectively.

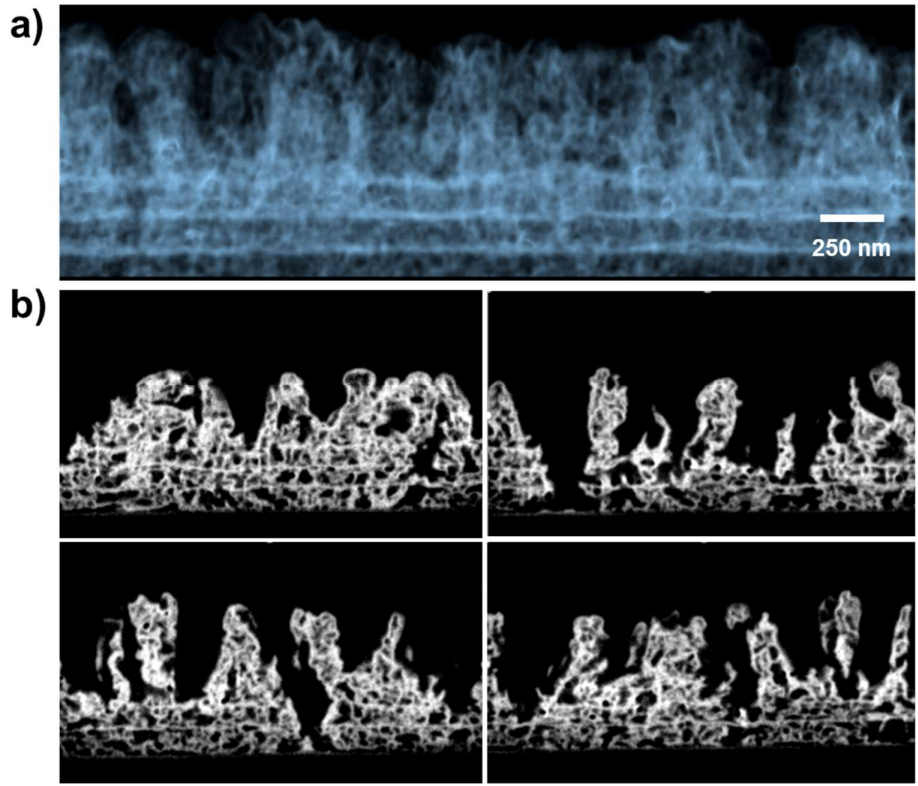

**Figure 3.- Determination of the pore structure by electron microscopy a)** FIB-SEM cross-sectional micrograph of a C5 porous $TiO_2$ sample. b) Sequence of FIB images from the previous sample showing the connectivity between the porous structures in the film.

An in-depth evaluation of the porous structure of these films can be obtained from the FIB-SEM analysis of the films. Figure 3 a) shows a FIB-3D micrograph of $TiO_2$ porous film deposited in 5 cycles. It is possible to distinguish in the micrograph the different interfaces corresponding to the first three deposition cycles reaching a thickness of around 500 nm. Apart from these three interfaces, the film is very porous and continuous. From cycle C4 and above, less interconnected porous columnar structures develop and the clear interface between cycles disappears. These columnar structures can also be recognized in the cross-sectional FESEM views of Figure 2 d)-e). The set of xz slices obtained by FIB etching at successive y positions shown in Figure 3 b) reveals the completely interconnected three-dimensional (3D) porosity and the lateral connection of the oxide nanostructures in the films (see the full video sequence of this FIB-SEM in S3 (Video S3)). The image sequence allows observing that porous structures are continuous and connect the film surface with the substrate. Apart from the interconnected porosity, a comparison between the FESEM and FIB-SEM micrographs reveals an extremely low-density 3D $TiO_2$ interconnected network in the films.

Figure 4a) shows the evolution of the film density (black dots) as a function of the layer thickness corresponding to a set of films prepared by 1 to 5 deposition/plasma oxidation cycles. The figure also shows the corresponding porosity values calculated using the relationship Γ = 100x(1-ρ/ρ$_r$) taking the density of compact titania as ρ$_r$= 4.23 g/cm$^3$.[54] The first deposition/plasma etching cycle yielded films with a density lower than 1 g/cm$^3$ and a porosity of 76.8%. As the number of cycles increases, the density decreases dramatically reaching a value of 0.47 g/cm$^3$ at 440 nm of thickness (C5) which corresponds to a film porosity as high as 88.4 %. These values increase slightly up to 88.8% when the thickness is increased up to 660 nm with another deposition cycle, reaching a minimum density of 0.45 g/cm$^3$. The shape of the density and porosity versus the layer thickness curves indicate that both magnitudes are not constant over the thickness. If we calculate the density/porosity profile (i.e., the density/porosity of each section of the film) as a function of film thickness (Figure 4 b)) it can be noted how the film density/porosity decreases/increases rapidly up to a thickness of about 220 nm where it reaches a value of 0.35 g/cm$^3$/91.8% and then decreases/increases more slowly until it reaches a minimum density of 0.29 g/cm$^3$ at 440 nm, which corresponds to the porosity of 93.1%. In the last studied cycle to reach 660 nm, the density/porosity changes the tendency and increases/decreases up to 0.40 g/cm$^3$/90.6%.

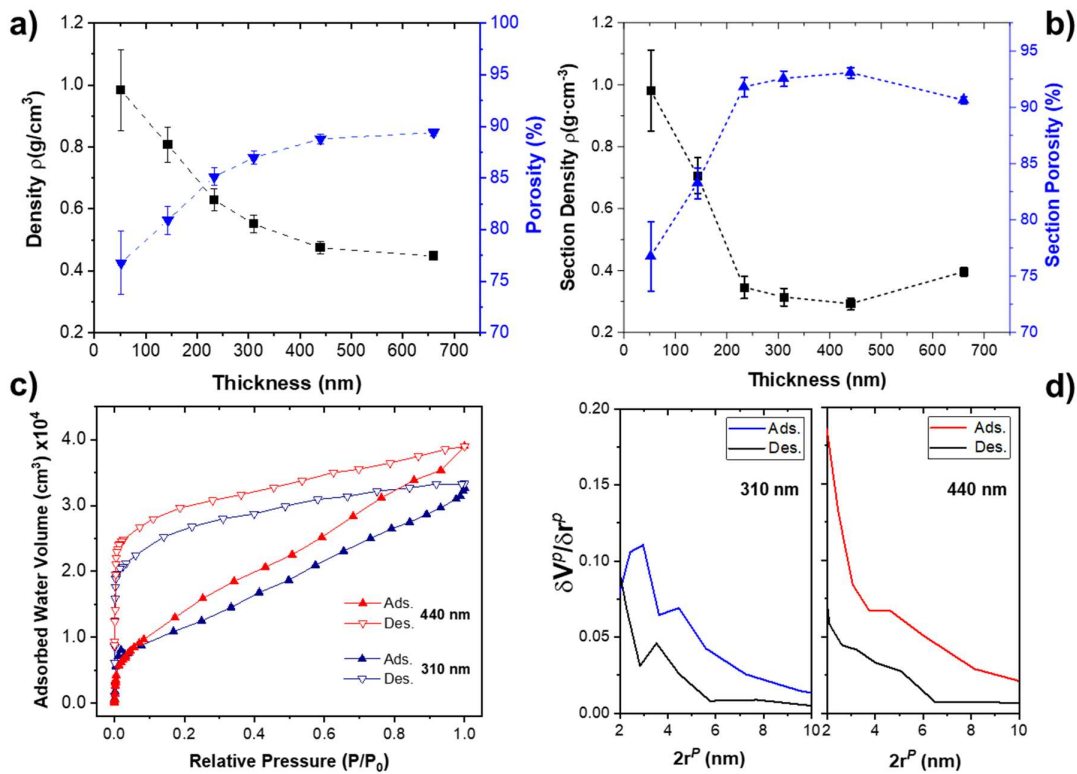

**Figure 4.- Characterization of the thin film density and porosity** a) Thin film overall density and porosity versus thickness. b) Evolution of the film density and porosity as a function of thickness. c) Normalized absorption-desorption water isotherms determined for films of 310 and 440 nm. d) Pores size distribution calculated from the adsorption isotherms in c) as indicated.

Thus, for thickness up to 200 nm the density and porosity values are in the range of those reported for aerogel films obtained by supercritical drying methods[55] and are even higher than typical values reported for TiO$_2$ aerogel films.[31,33,34,56] For this reason, we will hereafter refer to them as *aerogel-like* TiO$_2$ films, although their characteristic structure differs from those reported for conventional aerogels, as we will see in the next sections.

The porosity of the samples has been evaluated in detail by measuring water adsorption-desorption isotherms using a quartz crystal monitor. Figure 4c) shows normalized isotherms corresponding to two samples of 310 and 440 nm (C3 and C4 cycles). The two isotherms have similarly shaped adsorption-desorption curves. As expected, the total water adsorption capacity increases with film thickness. The isotherms belong to group IV of the IUPAC classification indicating that most of the porosity is due to mesopores[57] This type of curve presents an initial region that changes rapidly with adsorbed water volume and corresponds to the adsorption of micropores ($P/P_0 < 0.05$) and a second region that changes with a lower slope and corresponds to the adsorption of mesopores. The large absorption-desorption cycle hysteresis in both samples indicates a complex porous structure. However, unlike water adsorption isotherms reported for columnar porous TiO$_2$ thin films prepared by PECVD[28] and GLAD[49] determined by a similar methodology, the absorption-desorption cycles are fully reversible at room temperature even in the micropore region very likely due to a very high accessibility of the open porous structure of the films. Pore size distribution curves of the two studied films are shown in Figure 4b). The curves indicate that the two films have a relatively broad pore distribution in the range of 2 to 10 nm, with maxima in the region near 2 nm. The highest pore concentration corresponds to the values between 2-6 nm. In that region, the curves derived from the adsorption and desorption branches are similar.

*3.2 Synthesis of aerogel-like films at room temperature.*

As indicated in the experimental section, all the aerogel-like films shown in the figures above have been synthesized using deposition cycles at RT and plasma oxidation treatments at 120 °C. Oxidation at this mild temperature allows the oxidation of polymeric films up to ~200 nm thick range. Oxidation at room temperature, under the plasma conditions used in this work, is limited to thicknesses of less than ~50 nm. Above this thickness, only the surface of the polymeric layer is oxidized forming an oxide layer that protects the deeper polymeric layer from further oxidation (see Figure S4). However, such a limitation can be easily overcome by working with polymer sacrificial layers of reduced thicknesses (i.e., below 50 nm), making it possible to obtain fully oxidized porous structures at room temperature. Note this implies a higher number of cycles to obtain a given thickness in comparison with the previous examples (Figures 2-4). The synthesis at room temperature enhances the compatibility of the proposed method with temperature-sensitive substrates such as plastics and polymers. Figure 5 a) shows a TiO$_2$ aerogel-like film fully synthesized (deposition plus etching) at room temperature in 4 cycles. In a cross-sectional view (Figure 2 a), bottom), it is possible to observe a nanocolumnar structure delimited by gaps from the surface to the substrate. However, the very thin initial layer, in contact with the substrate, is denser with the development of the columnar structure some nm over the substrate. The corresponding HRTEM image (Figure 5b) shows that the film solid structures present a very low density where it is possible to identify continuous thin TiO$_2$ structures of no more than 4 nm thickness. These tenuous TiO$_2$ structures forming the solid part of the films are also interconnected. In summary, the structure of the RT-deposited films resembles those of the films prepared at mild temperature (Figure 2) with some differences in the scale of the observed features. Besides, the films are fully amorphous as shown in the digital electron diffraction pattern in Figure 5b).

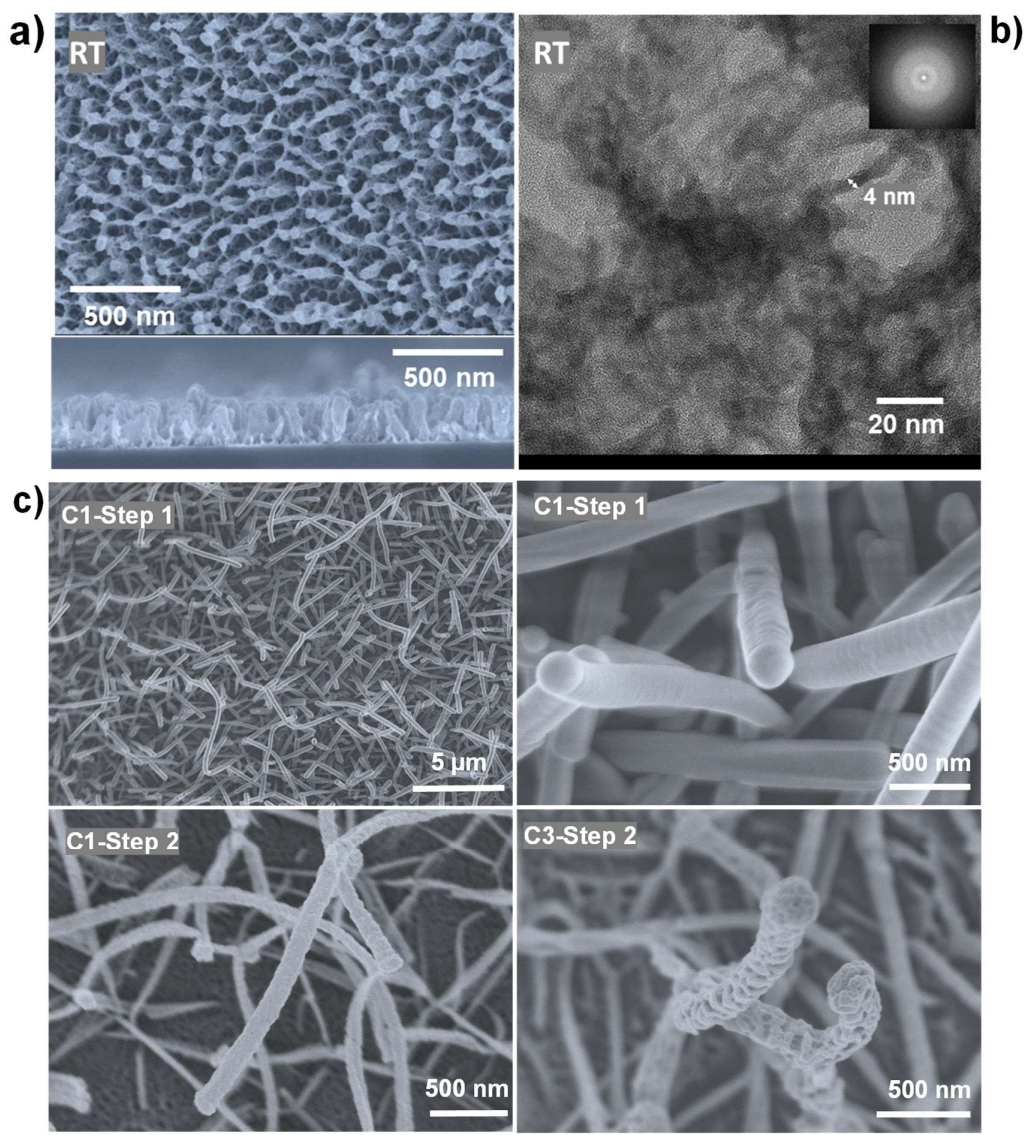

**Figure 5.- Examples of conformal coating of supported single-crystalline organic nanowires and deposition at room temperature** a) SEM normal and cross-sectional micrographs of a TiO$_2$ aerogel-like film fully synthesized (deposition plus etching) at room temperature in four deposition and oxidation cycles (C4). b) HRTEM micrograph of the film presented in a) with an inset showing the corresponding digital diffraction pattern. c) SEM micrographs showing examples of conformal deposition of RPAVD TiPc plasma polymer and TiO$_2$ aerogel-like films on single crystalline supported nanowires. RT stands for room temperature in panels (a)-(b) The key experimental steps are indicated in panel (c).

3.3 Conformal deposition.

One critical advantage of the RPAVD technique is that the plasma-assisted deposition gives rise to conformal coatings even on high-aspect-ratio supported nanostructures.[43] To demonstrate the extension of this critical advantage to the fabrication of aerogel-like materials, we have applied the combination of polymerization / oxidation cycles on supported single crystalline nanowires (ONWs). These nanowires are formed by self-assembly of conjugated small molecules by π-staking, see reference [45] for further experimental details. Figure 5 c) showcases characteristic SEM images of the different steps and cycles of the formation of the TiO$_2$ aerogel on a previously deposited array of ONWs. Due to the strong conformality of the RPAVD process, the plasma polymer (i.e., C1-Step 1) forms a uniform shell around the ONW with high homogeneity in thickness from the tip to the base of each nanowire (see reference [43]). After the etching step, the shell is converted forming the aerogel-like structure preventing the collapse of the 1D nanostructure but forming even smaller features in comparison with the thin film counterpart, as demonstrated in the two bottom panels in Figure 5 c). This result exemplifies the potential of the methodology for being applied to the coating of supported delicate nanostructures of different natures.

*3.4. Thin films crystallinity*

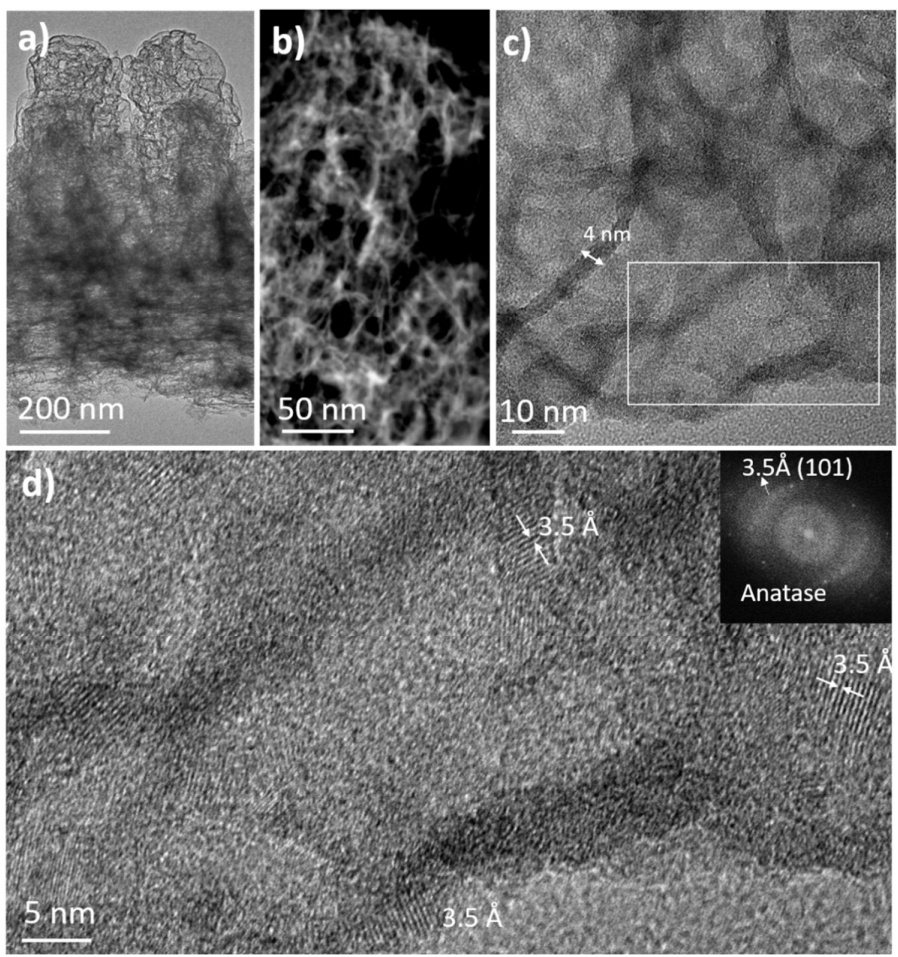

**Figure 6.- Determination of pore structure and embedded anatase nanocrystals by high-resolution electron microscopy.** STEM micrographs of an aerogel-like TiO$_2$ thin film prepared by five cycles of polymerization-plasma oxidation at 120 ºC: a) cross-sectional image, b) HAADF-STEM cross-sectional micrograph, c) High magnification TEM micrograph and d) HRTEM image of the square area marked in c) and the corresponding electron diffraction pattern.

The morphology and microstructure of the samples deposited by 5 cycles of polymerization/plasma oxidation can be observed in the STEM cross-sectional micrographs of Figure 6. The film presents an ultraporous structure (Figure 6a) that is more clearly observed in the HAADF-STEM image (Figure 6b). The very thin walls surrounding the pores can be neatly observed in the TEM images obtained at higher magnification (Figure 6c). This structure presents similarities to that observed in the RT sample (Figure 5b). In both cases, the samples are of very low density showing a similar sponge-like structure of percolated oxide walls of ~4 nm. Although both types of samples have the same internal structure despite the difference in thickness, there is a very noticeable difference between them that can be attributed to the difference in the plasma etching experimental conditions. While the films synthesized by plasma polymerization-oxidation cycles at room temperature are essentially amorphous (Figure 5b) the samples subjected to plasma oxidation at 120°C show crystalline domains of anatase nanoparticles distributed along their otherwise amorphous-like structure (see the HRTEM images and the digital diffraction patterns of Figure 6d).

As will be discussed in the next section, one of the applications of the $TiO_2$ ultraporous films is as selective contacts in perovskite solar cells. For this type of application, a higher crystalline film is necessary which is usually obtained by a thermal treatment at 450 °C during the cell manufacturing process. Figure 7 shows the TEM analysis of the film in Figure 6 after thermal annealing at 450 °C in air following the same protocol described for the preparation of $TiO_2$ electrodes in the experimental section. It is interesting to observe that the ultraporous structure remains stable after this treatment without significant modifications in the percolated structures which still maintain the same thickness and shape as in the original film (Figure 7 a)-c). However, the sample annealed at 450 °C presents a more crystalline structure, with a higher density of anatase domains in the range of 4-10nm. These domains are larger than those observed in the 120 °C sample (compare the HRTEM images and digital diffraction patterns of Figures 6d and 7d). Thus, a thermally activated diffusion process is responsible for the growth and fuse of the anatase nanocrystals within the amorphous matrix. It is worth highlighting that the growth of these anatase nanocrystals occurs without densification, cracking, or the collapse of the porous structure (Figure 7b), which is often a characteristic of thermally induced processes.[58] In our case, for the range of temperature studied it was not necessary to add a percentage of $SiO_2$ or to perform rapid heat treatments to preserve the porous structure as previously proposed to avoid thermal collapse in mesoporous $TiO_2$ films for high-temperature applications.[32,59]

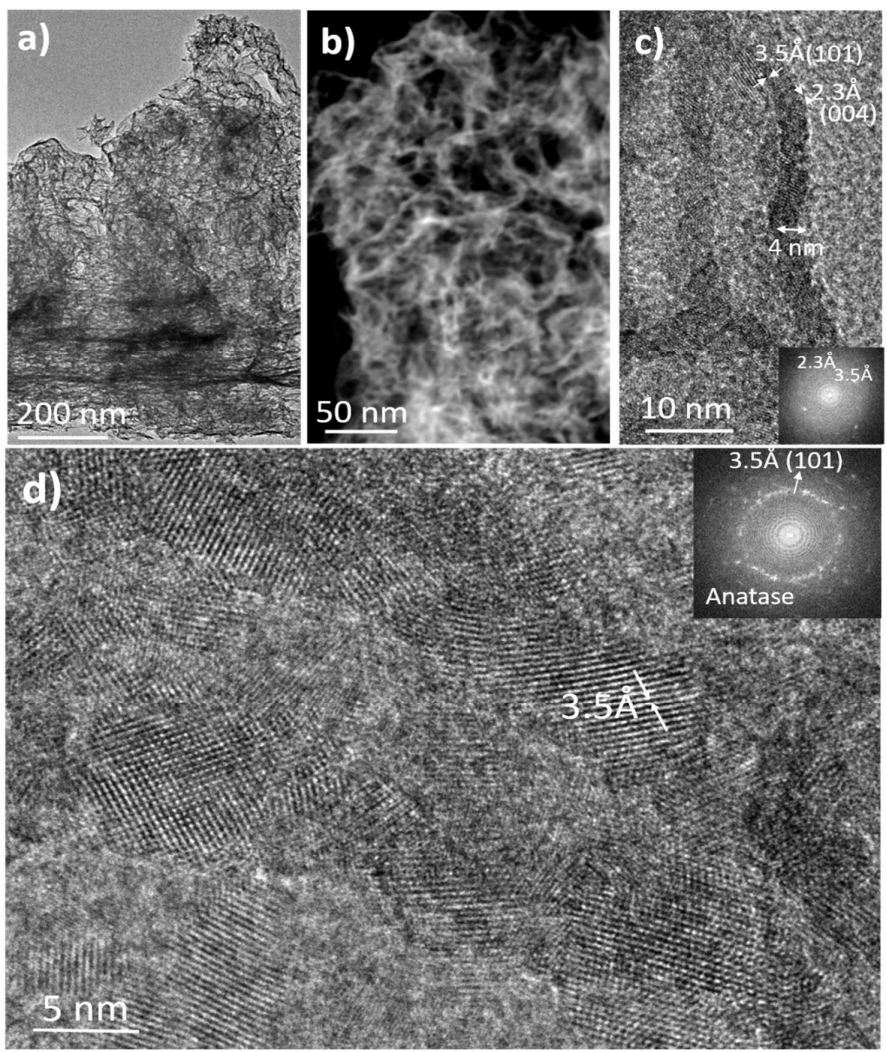

**Figure 7**.- **Embedded anatase nanocrystals and porous structure after annealing.** STEM micrographs of the aerogel-like film shown in Figure 6 after annealing at 450 °C in Ar atmosphere. a-c) Cross-sectional images at different magnifications. b) HAADF-STEM micrograph. d) HRTEM image and the corresponding digital diffraction pattern (inset).

*3.4 Optical properties*

The optical properties of aerogel-like $TiO_2$ films were studied on samples deposited in fused silica and stored in air with relative humidity between 45-50% for at least two months. Measurements were carried out at room temperature without any prior heat treatment. Figure 8a) shows the transmittance spectra of a set of samples from the first to the fifth cycle (C1-C5) deposited on fused silica, as well as an uncoated fused silica substrate. All the samples are transparent in the UV-VIS-NIR region without any significant absorption corresponding to the sacrificial plasma polymer film that is intensely absorbing in the visible as shown in Figure S1. The enlarged transmission region in Figure 6b) shows that all the films are antireflective with transmittances higher than the fused silica substrate. The overall transmittance of the films increases with the film thickness reaching values higher than 96% in the NIR region. Similarly, the reflectance of the films (Figure 8b)) decreases with the layer thicknesses reaching values in the range of 3.5-3.2% in the NIR. As expected, the overall light transmission of the films increases as the density of the films decreases with the thickness.

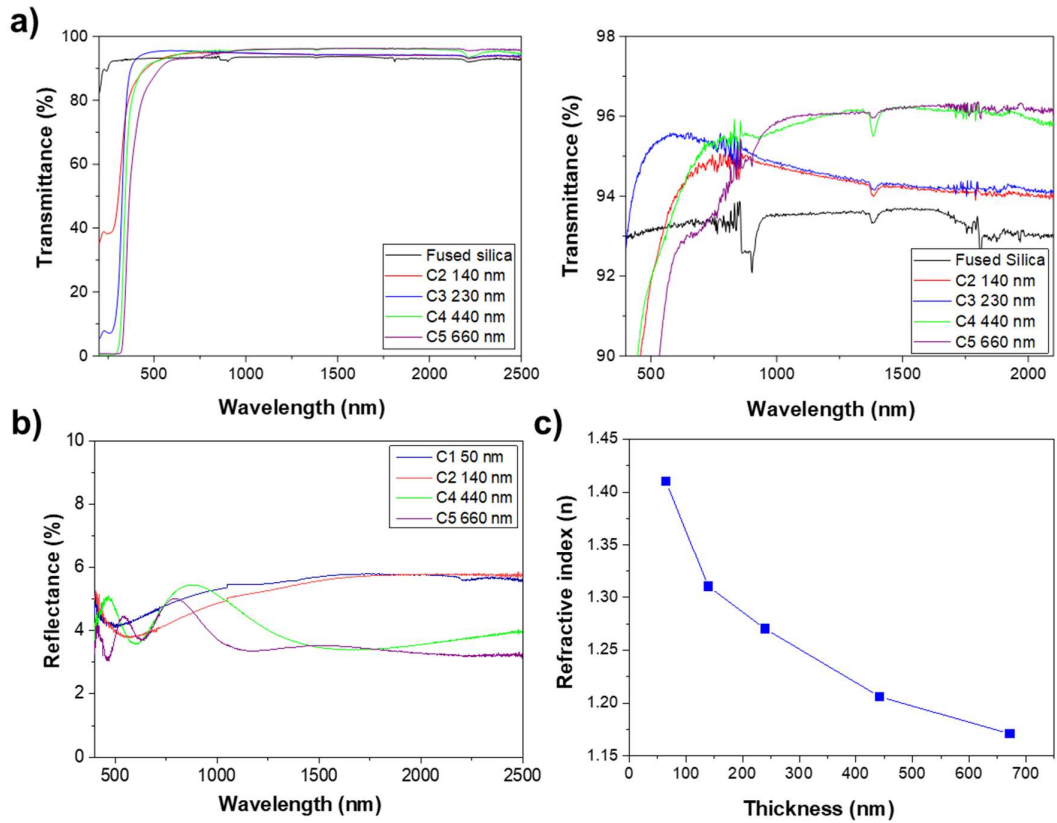

**Figure 8**.- a) Transmittance spectra in the UV-VIS-NIR region of a set of samples from C2 to C5, as well as an uncoated fused silica used as substrate. The figure at the right shows a zoom-in of the high transmittance zone. b) Reflectance spectra of a set of films for different cycles as labelled. c) Refractive index versus layer thickness determined at 735 nm by VASE.

Figure 8c) shows the evolution of the refractive index as a function of the thickness determined by variable angle spectroscopic ellipsometry. All the calculated values are lower than the fused silica refractive index (i.e., n=1.47 in the visible-NIR regions (i.e., n=1.47-1.43 in the spectral region 400-2500 nm) which explains the high light transmission values observed. The lowest refractive index value of 1.17 is achieved with the thickest sample. The optical characterization demonstrates the plasma-assisted deposition and etching methodology of aerogel-like films can find potential uses for the design of optical elements of antireflecting systems. The normal transmission and reflection values are lower than those reported for hierarchical mesoporous $TiO_2(SiO_2)$ films[57] and close to those reported for more complex optical coatings as moth-eyes structures obtained by relatively elaborated template methodologies.[30,60,61]

*3.5 Wetting properties*

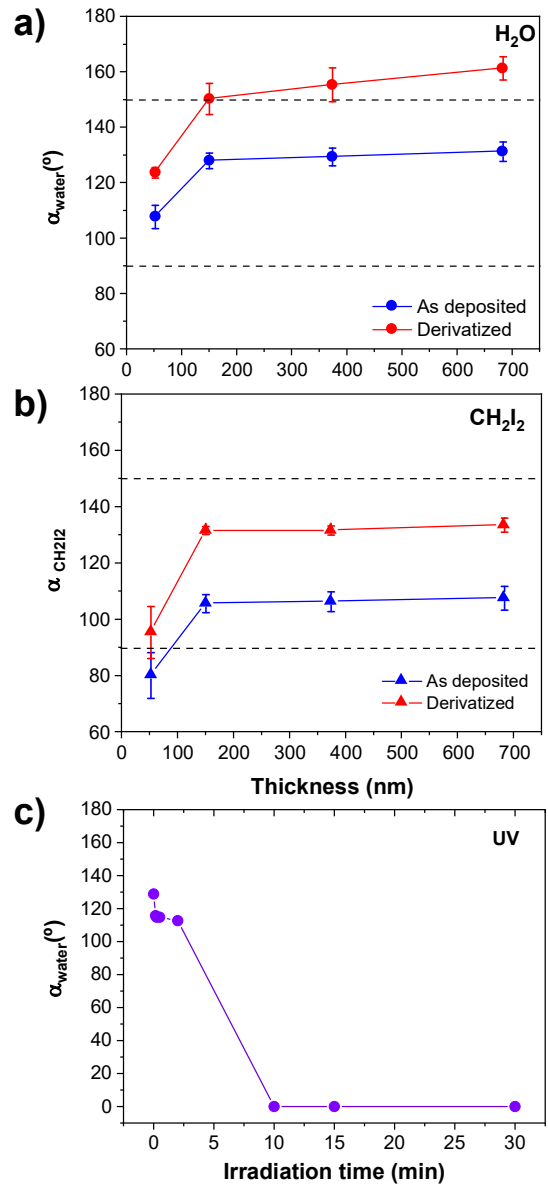

**Figure 9.-Wettability and photocatalytic properties of the TiO$_2$ aerogel-like films** a-b) Static wetting contact angle of aerogel-like films as a function of thickness for water (a) and diiodomethane (b). c) Water contact angle evolution under UV irradiation of a 440 nm thick TiO$_2$ aerogel-like film.

Figures 9a) and b) show in blue the evolution of the contact angle of water and diiodomethane droplets on the surface of the TiO$_2$ aerogel-like films, respectively. In the case of water, the aerogel-like film surfaces are hydrophobic over the whole range of thicknesses increasing from about 107 degrees for the 50 nm samples to 128 degrees for the 150 nm thick sample. At higher thicknesses, the values are stable increasing slightly up to ~131° for the thickest sample (i.e., 700 nm). Although some works have reported hydrophobicity values slightly higher than those reported here for rutile aerogel powder (c.a, 145°) [62] most of the authors commonly reported hydrophilicity of highly porous TiO$_2$ and TiO$_2$ aerogel surfaces.[59,63,64] Superhydrophobic water contact angle values have been reported mainly for TiO$_2$ aerogels chemically derivatized with non-hydrolyzable surface organic groups and silane derivatives.[62,65,66]

In the case of non-polar liquids such as diiodomethane, the evolution is similar: the thinner samples show a contact angle of ~80°, increasing up to ~106° for a thickness of 150 nm to remain stable in the range of 106-107° up to a thickness of 700 nm. Thus, the deposited pristine TiO$_2$ aerogel-like surfaces are omniphobic for thicknesses higher than 150 nm. These surfaces can be permanently chemically derivatized by attaching fluorinated chains to the surface hydroxyl groups by gas-phase silane chemistry (see experimental section). In this case, the contact angle of both water and diiodomethane increases over the entire range of thicknesses studied (blue lines in Figure 9(a)-(b)). The resulting derivatized films develop a more noticeable omniphobic behaviour throughout the thickness range studied. For water, superhydrophobicity values (i.e., θ>150°) are reached at thicknesses above 150 nm, reaching a contact angle above 160° for the thickest layers. The values corresponding to iodomethane also increase in parallel for all thicknesses, reaching remarkably high values above 130° in the range of 170-700 nm.

The omniphobic behaviour obtained before and after derivatization can be explained by the high roughness and porosity characteristic of the studied surfaces that allow reaching the so-called Cassie-Baxter state even with low surface tension liquids such as diiodomethane.[67] This state would be reached thanks to the re-entrant texture of the surface (see, for example, Figure 3) in which the liquid drop would not be able to wet the surface texture. Note that all the reported contact angle values correspond to porous layer surfaces deposited on flat substrates. Thus, the values shown could be optimized by introducing additional levels of surface structuring which is a common strategy for the design of liquid repellent hierarchical surfaces. Although these synthetic options will not be explored here, it is interesting to note the compatibility of the technique with conformal deposition on nanostructured substrates (see Figure 5 c). The stability of fluorinated TiO$_2$ aerogel surfaces has been tested up to three months after the derivatization process. Although further studies are needed to determine the stability over time under different environmental conditions, similar treatments have proven to be extremely robust.[51]

The hydrophobic and superhydrophobic aerogel-like films can become superhydrophilic by exposure to UV light for several minutes reaching a water contact angle of 0°, as shown in Figure 9c). These results demonstrate that the intrinsic photoactive nature of TiO$_2$ is preserved in the aerogel-like films. The effective absorption of water observed can be attributed to the presence of an interconnected porous network that allows water absorption by capillarity.[23,67,68] The possibility of producing highly hydrophilic TiO$_2$ surfaces by UV light activation is important for application as the development of perovskite solar cells which we will discuss in the next section.

*3.6 Aerogel-like films as selective electrodes in perovskite solar cells*

One important application of porous titanium oxide films is their use as an electron transport layer (ETL) in Graetzel solar cells and hybrid halide perovskite solar cells.[27,37] In fact, microstructural, electronic, and optical characteristics of the porous anatase ETL structure such as crystallinity, type of porosity, and sintered grain connectivity, as well as the level of doping, band-position, and transparency, are recognized as factors that determine the overall cell efficiencies in their different configurations.[69] During the last decade, hybrid metal-halide perovskite solar cells (PSCs) have emerged as a viable economic alternative to the current commercially available silicon photovoltaic technology.[70] The most efficient cells reported commonly use colloidal mesoporous TiO$_2$ as ETL in which the perovskite material is infiltrated by spin-coating.[22,44] Besides, TiO$_2$ aerogel films have been recently proposed as ETL in hybrid perovskite solar cells with promising results.[71,72]

Due to the open percolated porosity of the aerogel-like films presented in the previous section, they are good candidates as infiltration media for precursors in liquid solution. In this section, we have carried out a feasibility study about the preparation of perovskite cells using antireflective aerogel films as ETLs. For that, we have prepared perovskite solar devices following a well-established procedure[46] (see experimental section) in which RbCsMAFA perovskite [((FAPbI3)83(MAPbBr3)17 + 5% CsI) + 5% RbI] and Spiro-OMeTAD are the active layer and the hole transport layer (HTL), respectively.

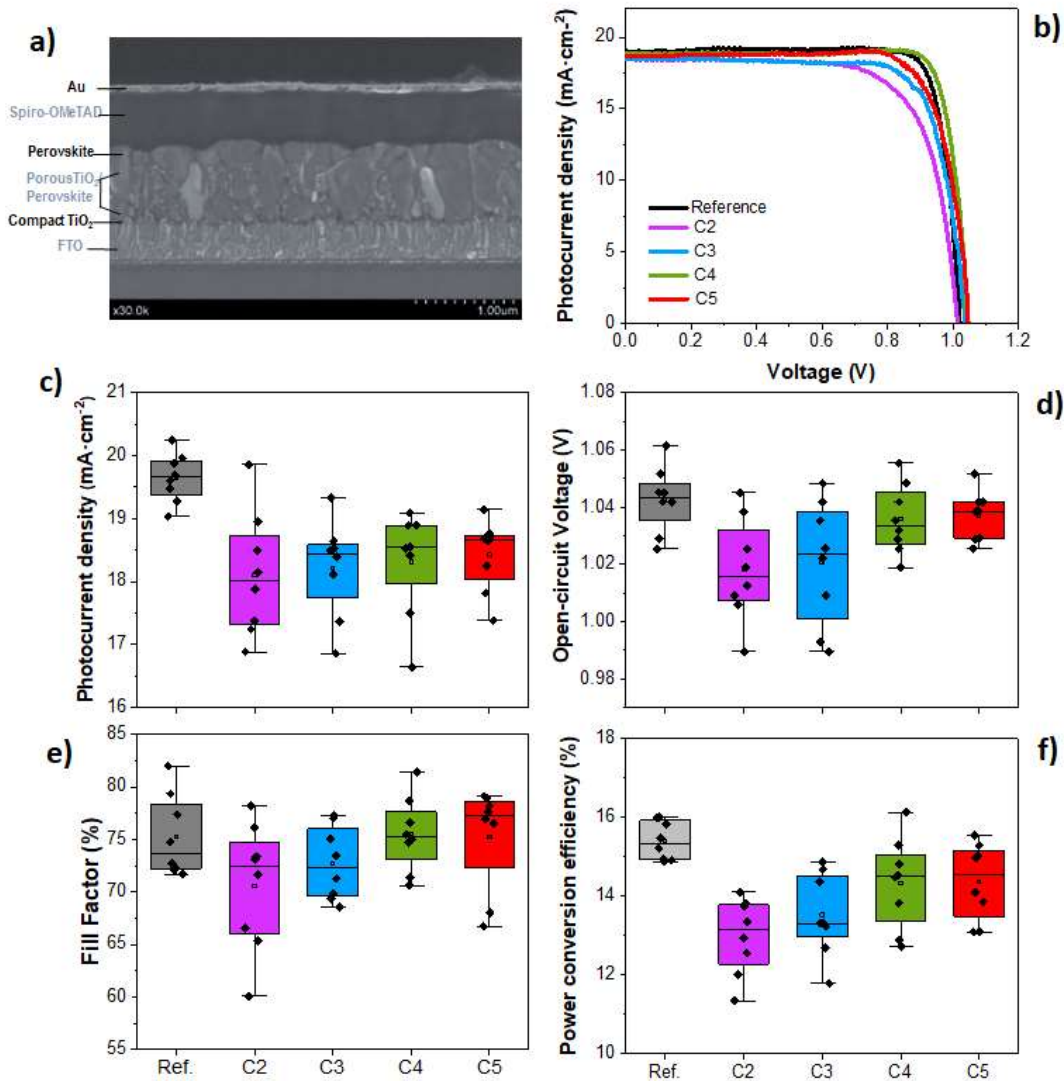

**Figure 10. Aerogel-like TiO$_2$ films as electron transport layers in perovskite solar cells.** a) SEM cross-sectional image of a hybrid perovskite solar cell fabricated on glass using aerogel-like TiO$_2$ films (C4) as ETL; b) Photocurrent-density vs Voltage curves measured under 1 sun-AM1.5G

illumination at a scan rate of 10mV/s and reverse scan of the champion set cells of every type. Photovoltaic data for reference samples are also included in the figure. Photovoltaic parameters obtained from the photocurrent-density curves: c) photocurrent density, d) photovoltage, e) fill factor, and f) power conversion efficiency.

Figure 10 a) shows a cross-sectional image of a complete PSC fabricated from an aerogel-like $TiO_2$ film of 4 cycles (C4) as ETL. As can be observed in the figure, the perovskite material is completely infiltrated into the porous film reaching the $TiO_2$ compact layer. Importantly, the perovskite material grows with well-defined and large grains (up to 800 nm) inside the $TiO_2$ aerogel-like scaffold. The photovoltaic response (current density-voltage curves, JV) of the champion solar devices using aerogel-like titania of different thicknesses as ETL is displayed in Figure 10 b). Figures 10c)-f) show the photovoltaic parameters obtained from these JV curves for at least 8 devices for the statistical analysis. Data from reference samples fabricated with commercial mesoporous titania are included for comparison. Note here that in the device synthesis procedure, both the commercial mesoporous and the aerogel-like $TiO_2$ thin films were heated up to 450 °C to increase their crystallinity. In all cases, a photovoltaic response is obtained (Figure 10b), i.e., the aerogel films can work as electron-selective electrodes despite their thickness. However, as can be seen in Figure 10c)-f), the photovoltaic parameters of the thinnest aerogel-like films (C2 and C3 cycles) are slightly below the reference devices. The reduced efficiency obtained for these devices is likely related to a lower photocurrent density and fill factor because of the higher series resistance (Fig. 10c)). On the other hand, the thicker aerogel-like films (C4 and C5 cycles) show efficiency values close to those of the reference devices. The most efficient cell of all the studied configurations, including the reference, corresponds to a C4 cell (PCE ~16.1%). It is widely reported in the literature that the mesoporous scaffold enhances the electron extraction from the perovskite layer due to decreased surface charge recombination rate leading to PSC of higher efficiency.[73,74] In this sense, the C4-C5 cycles would correspond to the optimal thickness region for efficient charge extraction by the aerogel-like film.

These results show that it is possible to develop perovskite cells with efficiencies similar to those achievable with commercial mesoporous $TiO_2$. Our findings also indicate that through proper optimization of the characteristics of the aerogel-like ETL (porosity, number of cycles, and thickness of each cycle) even more competitive efficiencies could be achieved. The methodology employed and the open porous structure of the aerogel-like films would make them compatible with a fully vacuum-based perovskite solar cell process.

*3.7 Generalization of the synthetic method.*

The XPS analysis showed that the remote plasma polymerized TiPc has an overall atomic composition similar to that of the molecule but with the notorious absence of chlorine removed as volatile species by the plasma interaction. Thus, the TiPc precursor ($C_{32}H_{16}Cl_2N_8Ti$) presents a relatively high C/Ti ratio of 32. Similarly, highly porous films can be obtained using the Titanyl Phthalocyanine precursor ($C_{32}H_{16}N_8OTi$) instead of TiPc by using the same synthetic methodology (see S6). However, the results differ significantly when using a different type of Ti-containing precursor, such as the Titanyl acetylacetonate ($C_{10}H_{14}O_5Ti$), which exhibits a much lower C/Ti ratio and includes 5 oxygen atoms bonded to Ti in the molecule. In this latter, the oxidation of the RPAVD polymer yields a relatively compact $TiO_2$ columnar structure (S6). We hypothesize that the high C/Ti ratio in the phthalocyanine precursor, and thus in the sacrificial

Ti-containing plasma polymer, is one of the key factors to obtain such high levels of porosity in the etched films. Other parameters determining the porous structure of the aerogel-like films are the number of deposition and etching cycles, and the thickness of the Ti-polymer layer in each cycle. Although the full possibilities of this synthetic approach have yet to be explored, the technique can be applied for the synthesis of not only aerogel-like films but also gradient index layers, porous multilayers, and porous complex core@shell nanoarchitectures. Importantly, the cycling plasma polymerization and plasma etching procedure can be applied to other metal-containing molecules to produce ultraporous oxide films. Two examples of highly porous $SiO_2$ and $Fe_2O_3$ thin films produced by the same methodology from Si(IV) and Fe(II) phthalocyanines respectively (see Supplementary Section S6).

**4.-Summary and conclusions**

In this work, we present a plasma-enabled deposition approach for the fabrication of aerogel-like oxide thin films by a combined plasma polymerization and oxygen plasma etching sequential process of a Ti(IV) phthalocyanine precursor. The plasma polymerization process is carried out by remote plasma-assisted vacuum deposition to give rise to homogeneous and crosslinked films with an overall stoichiometry very similar to that of the precursor molecule. The Ti polymer films are then used as precursors for the formation of $TiO_2$ porous films. The pore development is produced by the carbon and nitrogen removal of the Ti-containing plasma polymer by the formation of volatile species due to the interaction with the energetic oxygen species of the plasma. By adding successive polymerization and plasma etching steps, it has been possible to increase the overall porosity of the resulting oxide and control the thickness as well as the optical and surface properties of the films. One of the most interesting aspects is that by employing this method it is possible to obtain $TiO_2$ films with extremely low densities reaching values characteristic of aerogel materials, i.e. overall porosity values close to 90%.

The physicochemical characteristics of the oxide (i.e., stability of the oxidation states, crystallinity, thermal properties, among others) are also factors determining the properties of the ultraporous films obtained that will be studied in future works. In the case of the $TiO_2$ the as-deposited aerogel-like films are antireflective in the VIS and NIR range, omniphobic for thicknesses higher than 150 nm, and photoactive under UV illumination. The films deposited using etching treatments at room temperature are amorphous but those etched at 120 °C are partly crystalline. More importantly, the films can be annealed in air at 450 °C to increase the crystalline fraction without observable significant changes in the porous structure. The aerogel-like presents an open percolated porosity that is accessible to liquids and gases. A feasibility study about the application of aerogel-like films as electron transport layers in perovskite solar cells has been carried out thanks to the possibility of infiltrating the porous structure with the perovskite liquid precursors. The results show that the solar cell performance can be improved by optimizing the ETL porous layer to achieve values close to those of reference cells fabricated using commercial mesoporous anatase. Our results indicate that the process can potentially be a simple alternative for the synthesis of porous ETLs that can be combined with the vacuum synthesis of the different components of a perovskite cell.

It is important to mention that alternative solid precursors can be employed to control the pore size distribution of the aerogel films.  As demonstrated, the method is straightforwardly extendable to other metal complexes to yield porous layers of metal oxide with controlled stoichiometry. In addition to their general character from the point of view of the layer composition, it is worth stressing relevant advantages from the sustainability point of view such as the solventless nature of the process, along with the use of non-toxic, non-flammable and highly environmentally friendly solid precursors. Indeed, the proposed synthetic route employing vacuum processable precursors that are sublimated from punctual Knudsen cells ensures the efficient utilization of the entire precursor on the substrates. Consequently, the presented method offers a universal procedure for the synthesis of conformal aerogel-like thin films by vacuum and plasma processes scalable to industrial production, which may pave the way for the development of novel optoelectronic and photonics devices. Furthermore, the reported synthetic methodology opens exciting possibilities in areas such as biomaterials and catalytic supports.

Acknowledgements

We thank the projects PID2019-110430GB-C21, PID2019-110430GB-C22, TED2021-130916B-I00, and PID2022-143120OB-I00 funded by MCIN/AEI/10.13039/501100011033 and by "ERDF (FEDER) A way of making Europe, Fondos NextgenerationEU and Plan de Recuperación, Transformación y Resiliencia", and the Consejería de Economía, Conocimiento, Empresas y Universidad de la Junta de Andalucía (PAIDI-2020 through projects P18-RT-3480 and US-1381057). FJA thanks the EMERGIA Junta de Andalucia program. FJF thanks the University of Seville (VI PPIT-US). The project leading to this article has received funding from the EU H2020 program under grant agreement 851929 (ERC Starting Grant 3DScavengers).

# Supplementary Information

# Conformal TiO$_2$ aerogel-like films by plasma deposition: from omniphobic antireflective coatings to perovskite solar cells photoelectrodes

Jose M. Obrero[a], Lidia Contreras-Bernal[a], Francisco J. Aparicio,[a,b] Teresa C. Rojas[a], Francisco J. Ferrer,[c] Noe Orozco,[a] Zineb Saghi,[d] Triana Czermak,[a] Jose M. Pedrosa,[e] Carmen López-Santos,[a,b] Kostya (Ken) Ostrikov,[f] Ana Borras,[a] Juan Ramón Sánchez-Valencia,[a]* Angel Barranco.[a]*

a) Nanotechnology on Surfaces and Plasma Laboratory, Materials Science Institute of Seville (CSIC-US), C/ Américo Vespucio 49, 41092, Seville, Spain.

b) Departamento de Física Aplicada I, Escuela Politécnica Superior, Universidad de Sevilla, Spain. c/ Virgen de Africa 4101

c) Centro Nacional de Aceleradores (CNA, CSIC-Universidad de Sevilla)

d) Univ. Grenoble Alpes, CEA, LETI, F-38000 Grenoble, France

e) Departamento de Sistemas Físicos, Químicos y Naturales. Universidad Pablo de Olavide, Ctra. Utrera Km. 1, 41013 Sevilla, Spain.

f) School of Chemistry and Physics and Centre for Materials Science, Queensland University of Technology (QUT), Brisbane, QLD 4000, Australia.

corresponding authors: angel.barranco@csic.es, jrsanchez@icmse.csic.es

**S1.- UV-Vis characterization of a TiPC plasma polymer**

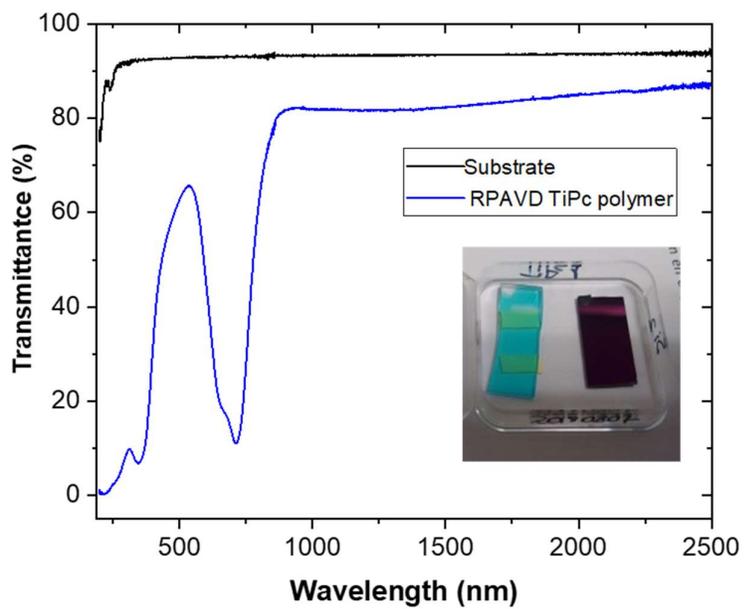

**Figure S1.- UV-VIS-NIR spectrum of a TiPc plasma polymer.** UV-VIS-NIR spectra of a RPAVD TiPc plasma polymer a as well as the bare fused silica substrate used. The inset show a picture of the TiPc plasma polymer deposited on fused silica and Si(100).

**S2.- XPS analysis**

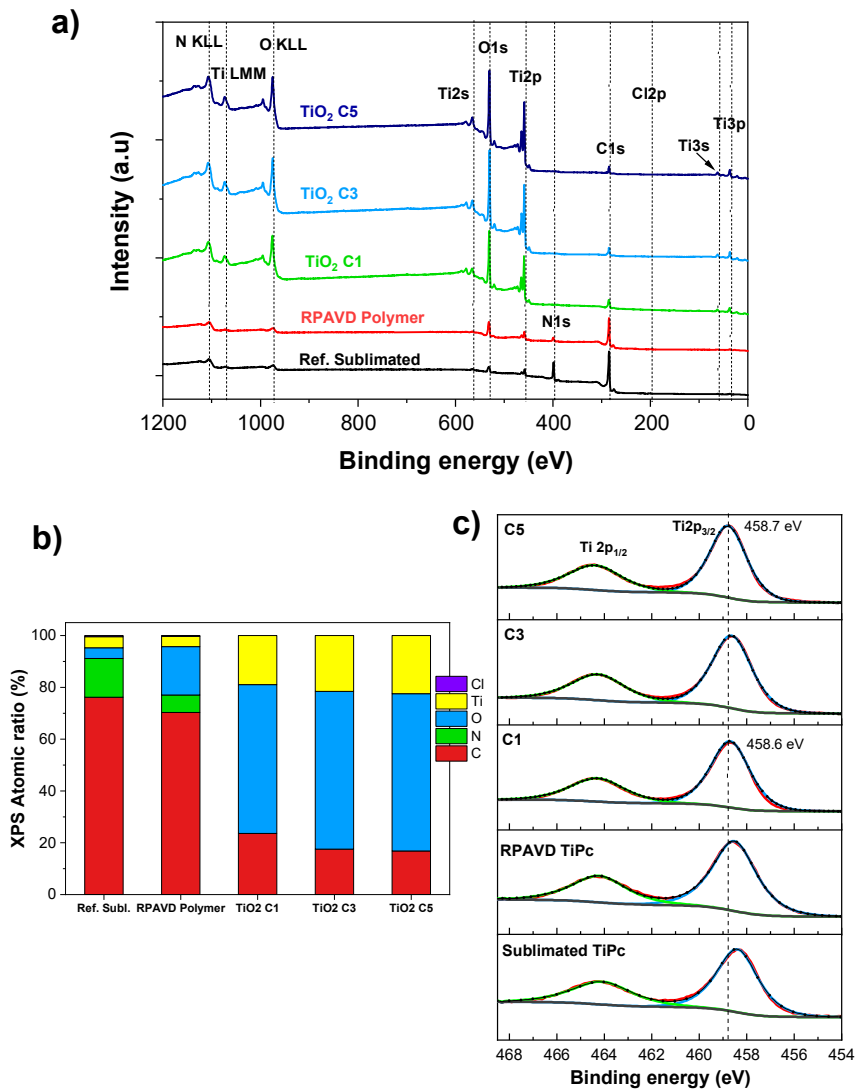

**Figure S2. XPS characterization** a) XPS survey spectra of sublimated, polymerized and three aerogel-like TiO2 thin films as indicated. b) Relative atomic ratios deduced from the XPS analysis of the samples in a). c) XPS spectra of the Ti2p core level of the samples in a). The Ti2p3/2 binding energies of the porous TiO2 samples are in the range 450.7-458.6 characteristics of $TiO_2$ (NIST). Note that the TiPc plasma polymer and sublimated TiPc reference are slightly shifted to lower binding energies due to the different chemical environment of the cation although also corresponds to Ti(IV) core levels. All the XPS analyses correspond to as-deposited samples without any surface cleaning procedure.

**S3.- FIB-SEM characterization**

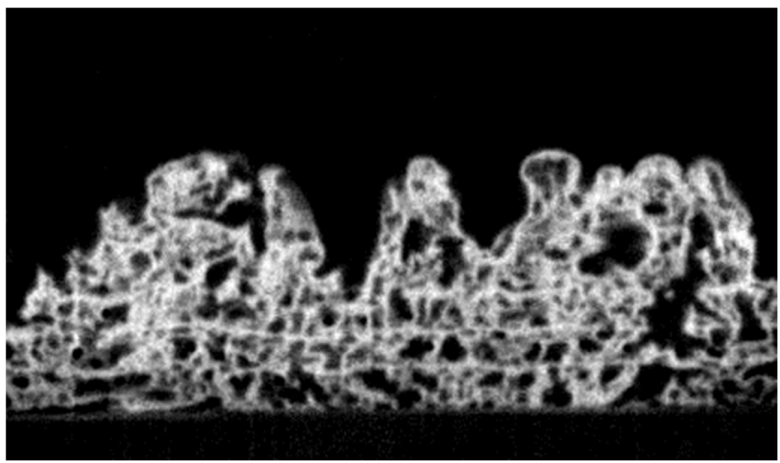

**Figure S3**.- **Aerogel-like film porous structure characterization**. FIB-SEM characterization of a aerogel-like film. Image taken from the video sequence of the FIB characterization of an aerogel-like TiO$_2$ thin film. The full video corresponds to the sample in Figure 3 (see main text). This video is attached as Video S1 in Supporting information material.

**S4.- Plasma etching at RT**

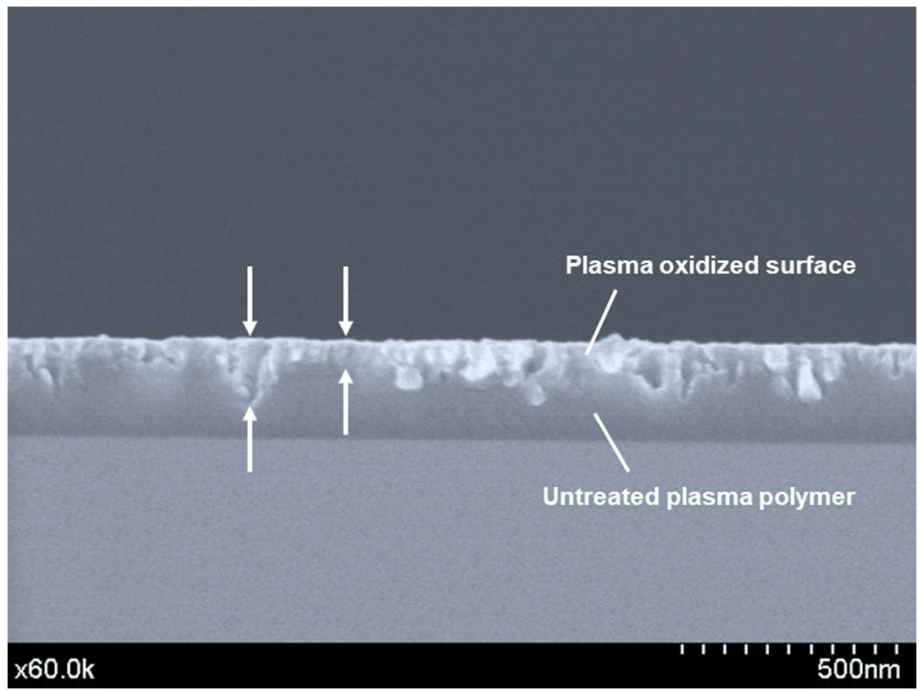

**Figure S4.- Plasma etching treatment at room temperature.** Cross-sectional SEM micrograph of a ~210 nm thick RPAVD TiPc plasma polymer subjected to an oxygen plasma etching treatment at room temperature as described in the experimental section. The image shows how the oxidation of the plasma polymer film is incomplete and restricted to a surface region of ~50-100 nm thick as indicated.

**S5.- XRD**

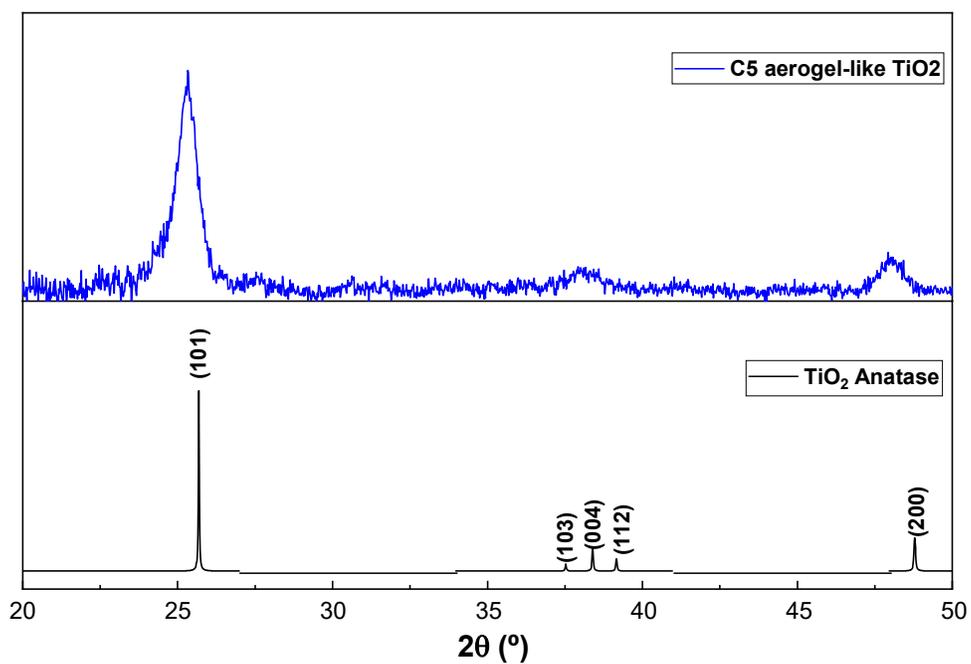

**Figure S5.- XRD characterization of a partially crystalline aerogel-like thin film.** Glancing angle XRD of a C5 of 640 nm thick aerogel-like $TiO_2$ thin film. Anatase phase diffraction peaks are included for comparison (catalog JCPDS-ICDD 2003 file number 78-2486)

**S6. Examples of synthesis of porous oxide films using additional metal containing precursors.**

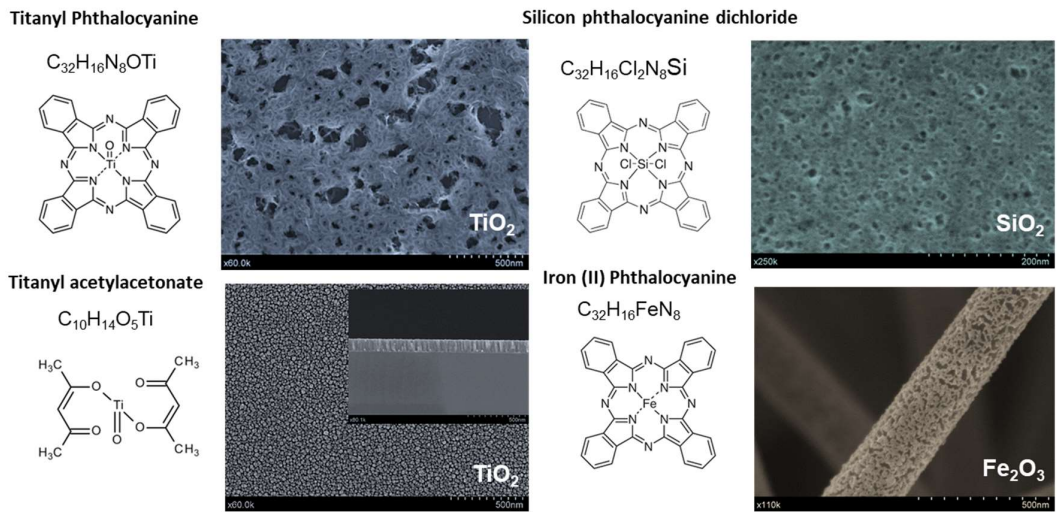

**Figure S6.- Generalization of the aerogel-like thin film synthetic procedure**. Examples of porous films obtained through the sequence of plasma polymerization and plasma etching described in the text but utilizing different metal-containing precursors. The three phthalocyanine precursors yield highly porous films, displaying distinct similarities to the aerogel films discussed in the text. However, the polymerization and plasma oxidation of titanyl acetylacetonate result in a more compact and denser microstructure characterized by packed nanocolumns. A comprehensive investigation of the properties of these films falls beyond the scope of this study.